\documentstyle[aps]{revtex}

\begin{document}
\author{M.S. Hussein$^{1}$, M.P. Pato$^{1,2}$, and J.C. Wells$^{2,3\thanks{%
Present address: Center for Computional Sciences, Oak Ridge National
Laboratory, Oak Ridge, Tenessee, 37831-6373, U.S.A.}}$}
\address{$^1$Instituto de F\'{\i}sica, Universidade de S\~{a}o Paulo, \\
C.P.66318, 05315-970, S\~{a}o Paulo, SP, Brazil }
\address{$^2$Institute for Theoretical Physics, University of California, \\
Santa Barbara, 93106-4030, USA }
\address{$^3$Institute for Theoretical Atomic and Molecular Physics, \\
Harvard-Smithsonian Center for Astrophysics,\\
Cambridge, Massachusetts, 02138, USA}
\title{Causal Classical Theory of Radiation Damping}
\maketitle

\begin{abstract}
It is shown how initial conditions can be appropriately defined for the
integration of Lorentz-Dirac equations of motion. The integration is
performed {\it forward} in time. The theory is applied to the case of the
motion of an electron in an intense laser pulse, relevant to nonlinear
Compton scattering. 
\end{abstract}

The advent of a new generation of extremely high power lasers that uses
chirped pulse amplification has put into focus the classical description of
the dynamics of relativistic electrons. Under the action of high intensity
electromagnetic field, a major ingredient of the dynamics is the electron
self-interaction which implies in the damping of the movement caused by the
interaction of the charge with its own field. The derivation of the damping
force has been reviewed recently\cite{Luhm} revealing its relativistic
origin asociated to the asymmetry introduced by the Doppler effect in the
forward and backward emission of radiation. The inclusion of this force in
the equation of motion leads to the nonlinear covariant Lorentz-Dirac(LD)
equation\cite{Dirac} for a point charge.

Nonlinear effects have been observed in recent experiments with
intense-laser relativistic-electron scattering at laser frequencies and
field strengths where radiation reaction forces begin to become significant%
\cite{Hart,Bula}. Relativistic nonlinear Thompson scattering has also been
observed\cite{Yuan}. These experiments justify the recent attention the
classical electrodynamics theory(CED) has received. A self-consistent
classical theoretical treatment of the radiative reaction force would also
be useful in simulating future electron accelerators\cite{HP92}. Further,
the kind of study reported upon in this paper may be useful in the quantal
treatment\cite{FLow}.

A certain number of problems, conceptual and practical, is known to be
associated with LD equation. These difficulties may be traced to the fact
that it contains a dependence on the derivative of the acceleration which
implies in the necessity of imposing, in order to solve it, an extra
condition besides the usual initial conditions on position and velocity of
classical mechanics. It has been estabilished by Rohrlich that this
condition is given by the asymptotic constraint that at the far future, when
the fields vanish, the acceleration should also vanish. The solution which
is obtained however, when extrapolated to the instant when the external
force is applied violates causality. On the other hand, from the practical
point of view, a condition put in the far future is awkward to be
implemented specially in a scattering situation.

A recent book\cite{Yagh} has been devoted to this question of causality
violation. A clear explanation to its origin was given and it has been shown
how it can be solved in the context of the CED. On the other side, attempts
has been done in order to replace Rohrlich condition by an equivalent
initial condition\cite{Agui,Vill}. The purpose of this letter is to apply
these modern advances, to the case of the classical description based in LD
equation, of the movement of an electron interacting with a short, strong
laser pulse. This problem has been recently discussed in the literature
Ref.\ \cite{Hart}, which provided numerical solutions of the Lorentz-Dirac
equation. The integration was performed backwards in time so that the
unphysical, exponentially growing homogeneous solutions of LD would damp
out, resulting in a numerical stable solution.

We are going to show that Lorentz-Dirac equation of motion can be integrated
forward in time with conditions specified at $t=0$. The idea is to construct
the series solution of LD equation. The initial acceleration is then
provided by replacing in the series the velocity by their initial value at
the instant when the external force is applied. It is easy to show that the
solution obtained with this procedure, when extrapolated to the distant
future, satisfies Rohrlich condition. However, we still have to cope with
the existence of the unphysical runaway solutions which although formally
eliminated troubles the process of numerical integration. Then, by combining
the recursive use of the series solution with implicit methods of numerical
integration we show that the process of integration forward in time can be
performed.

We write Lorentz-Dirac (LD) equation of motion as
\begin{equation}
F_{\mu }^{ext}=a_{\mu }-\epsilon \left( \frac{d^{2}v_{\mu }}{d\tau ^{2}}%
+v_{\mu }a_{\lambda }a^{\lambda }\right)   \label{2}
\end{equation}
where $v_{\mu }$,$a_{\mu }$ and $F_{\mu }^{ext}$ are, respectively, the
four-vector components of the velocity,acceleration and of the external
force given explicitly by

\begin{equation}
v_{\mu }=\gamma \left( 1,{\bf \beta }\right) ,  \label{4}
\end{equation}

\begin{equation}
v^{\mu }=\gamma \left( 1,-{\bf \beta }\right) ,  \label{4a}
\end{equation}

\[
a_\mu =\frac{dv_\mu }{d\tau } 
\]

and

\begin{equation}
F_{\mu }^{ext}=\gamma \left( {\bf \beta }\cdot {\bf F}_{ext},{\bf F}%
_{ext}\right) .  \label{6}
\end{equation}
In these equations, $\tau $ is the dimensionless proper time $d\tau =\omega
_{0}dt/\gamma $, $\gamma $ is the relativistic factor $\gamma =\frac{1}{%
\sqrt{1-\beta ^{2}}}$ with ${\bf \beta }=\frac{1}{c}\frac{d{\bf r}}{dt}$ and 
$\omega _{0}$ is the frequency of the laser pulse with which the electron is
interacting. We follow here the same units of Ref.~\cite{Hart}. With these
definitions, it can be easily verified that $v_{\mu }a^{\mu }=0$ as it
should. The quantity $\varepsilon =\omega _{0}\tau _{0},$ where $\tau
_{0}=e^{2}/m_{e}c^{3}=0.626\times 10^{-23}s$ is the Compton time scale. We
are assuming that the external force applies at $\tau =0$ and the second
term in the right hand side of the equation represents the damping force
arising as the charge starts to radiate\cite{Rohr}.

By contracting the Lorentz-Dirac equation with $a_{\mu }$, it is found that $%
\left( a_{\mu }a^{\mu }\right) \left( \tau \right) =C\exp (2\tau /\epsilon )$
is a solution of the resulting homogeneous equation (no external force) for
times greater than $\tau _{0}$. These are the so-called runaway solutions.
They are eliminated by imposing the Dirac-Rohrlich (DR) condition

\begin{equation}
\lim\limits_{\tau \rightarrow \infty }a_\mu =0  \label{8}
\end{equation}

when

\begin{equation}
\lim\limits_{\tau \rightarrow \infty }F_\mu ^{ext}\left( \tau \right) =0.
\label{10}
\end{equation}

In contrast to the usual initial value problem encountered in classical
physics, where all quantities, position, velocity, are fixed at $t=0$, the
above asymptotic condition, Eq. (\ref{8}) recast the problem into a boundary
value one where the $x,v$ are known at $\tau =0$ and $a$ is forced to be
zero at $\tau =\infty $, in accordance with the condition on the force, Eq. (%
\ref{10}). A particular situation is that of Ref.\ \cite{Hart}, in which the
problem was solved in the rest frame of the particle at some future time
with three final homogeneous conditions. In both situations however, the
numerical integration has to be performed backwards in time. Although this
procedure eliminates the unphysical runway solutions, it is uncomfortable,
from the point of view of applications, to have a condition given at some
final time. It is therefore desirable to have an equivalent condition on the
acceleration defined at the initial time.

For an electron moving in an electromagnetic field the external force is
given by 
\begin{equation}
F_{\mu }^{ext}=-(\partial _{\nu }A_{\mu }-\partial _{\mu }A_{\nu })v^{\nu }
\label{12}
\end{equation}
where the quantity $A_{\mu }$ is the vector potential given in units of $%
m_{0}c/e$. For a linearly polarized laser pulse, 
\begin{equation}
A_{\mu }\equiv (\Phi /c,{\bf A}),\;\;\;\;{\bf A}=\hat{x}A_{x}(\phi
),\;\;\;\;\Phi =0  \label{16}
\end{equation}
which is a function of the invariant phase of the traveling wave, 
\begin{equation}
\phi =k^{\mu }x_{\mu }(\tau )=x_{0}-z\;,  \label{20}
\end{equation}
where $k^{\mu }=(1,0,0,1)$ is the dimensionless laser wave number. 

Following Ref.\ \cite{Hart} we use $\phi $ as the independent variable to
recast the Dirac Lorentz equation as 
\begin{eqnarray}
\frac{d^{2}v_{x}}{d\phi ^{2}} &=&v_{x}\left( \left( \frac{d{\bf v}}{d\phi }%
\right) ^{2}-\left( \frac{d\gamma }{d\phi }\right) ^{2}\right) +\frac{1}{u}%
\left( \frac{dv_{x}}{d\phi }\left( \frac{1}{\varepsilon }-\frac{du}{d\phi }%
\right) +\frac{A}{\varepsilon }G(\phi )sin\phi \right)   \label{28} \\
\frac{d^{2}v_{z}}{d\phi ^{2}} &=&v_{z}\left( \left( \frac{d{\bf v}}{d\phi }%
\right) ^{2}-\left( \frac{d\gamma }{d\phi }\right) ^{2}\right) +\frac{1}{u}%
\left( \frac{dv_{z}}{d\phi }\left( \frac{1}{\varepsilon }-\frac{du}{d\phi }%
\right) +\frac{v_{x}}{u}\frac{A}{\varepsilon }G(\phi )sin\phi \right) 
\label{30}
\end{eqnarray}
where the laser pulse electric field 
\begin{equation}
\frac{dA_{x}(\phi )}{d\phi }=AG(\phi )sin\phi :,  \label{32}
\end{equation}
has been introduced, with $A$ being the maximum amplitude of the pulse, $%
G(\phi )=e^{-(\phi /\Delta \phi )^{2}}$ is a unit Gaussian envelope of width 
$\Delta \phi $. Note that 
\begin{equation}
\left( \frac{d{\bf v}}{d\phi }\right) ^{2}=\left( \frac{dv_{x}}{d\phi }%
\right) ^{2}+\left( \frac{dv_{z}}{d\phi }\right) ^{2}\;,  \label{33a}
\end{equation}

\begin{equation}
\frac{d\gamma }{d\phi }=\frac{v_{x}}{\gamma }\frac{dv_{x}}{d\phi }+\frac{%
v_{x}}{\gamma }\frac{dv_{z}}{d\phi }  \label{33b}
\end{equation}
and

\begin{equation}
\frac{du}{d\phi }=\frac{d\gamma }{d\phi }-\frac{dv_{z}}{d\phi }.  \label{33c}
\end{equation}

Substituting these relations into Eqs.\ (\ref{28} ) and (\ref{30}) we find 
\begin{eqnarray}
\frac{d^{2}v_{x}}{d\phi ^{2}} &=&v_{x}Q+\frac{1}{u}\left( \frac{1}{%
\varepsilon }\frac{dv_{x}}{d\phi }-\frac{v_{x}}{\gamma }\left( \frac{dv_{x}}{%
d\phi }\right) ^{2}+\left( 1-\frac{v_{z}}{\gamma }\right) \frac{dv_{x}}{%
d\phi }\frac{dv_{z}}{d\phi }+\frac{A}{\varepsilon }G(\phi )sin\phi \right)
\label{34} \\
\frac{d^{2}v_{z}}{d\phi ^{2}} &=&v_{z}Q+\frac{1}{u}\left( \frac{1}{%
\varepsilon }\frac{dv_{z}}{d\phi }-\frac{v_{x}}{\gamma }\frac{dv_{x}}{d\phi }%
\frac{dv_{z}}{d\phi }+\left( 1-\frac{v_{z}}{\gamma }\right) \left( \frac{%
dv_{z}}{d\phi }\right) ^{2}+\frac{v_{x}}{u}\frac{A}{\varepsilon }G(\phi
)sin\phi \right)  \label{36}
\end{eqnarray}
where

\begin{equation}
Q=\left( \frac{dv_{x}}{d\phi }\right) ^{2}\left( 1-\frac{v_{x}^{2}}{\gamma
^{2}}\right) +\left( \frac{dv_{z}}{d\phi }\right) ^{2}\left( 1-\frac{%
v_{z}^{2}}{\gamma ^{2}}\right) -\frac{2v_{x}v_{z}}{\gamma ^{2}}\frac{dv_{x}}{%
d\phi }\frac{dv_{z}}{d\phi }  \label{40}
\end{equation}

and 
\begin{eqnarray}
\gamma &=&\sqrt{1+v_{x}^{2}+v_{z}^{2}}  \label{44} \\
u &=&\gamma -u_{z}\;.  \label{46}
\end{eqnarray}

In Ref.\ \cite{Hart}, by specifying final homogeneous conditions on the
acceleration and the velocity and then integrating {\it backward} in time,
the solution to these equations were obtained at all times. We now want to
show that this problem can also be solved by specifying initial conditions
on the motion and integrating {\it forward} in time. 

For the ``initial'' velocity of our method, we use the final velocity of the
backward integration method of \cite{Hart}. As for the initial acceleration
we employ the first terms of the series generated by expanding the equations
of motion, in terms of the small quantity $\varepsilon =\omega _{0}\tau _{0}$%
. To obtain this series we write the two components of the equations of
motion (\ref{34}) and (\ref{36}) as 

\begin{eqnarray}
\frac{dv_{x}}{d\phi } &=&-AG(\phi )sin\phi +\varepsilon \left[ u\left( \frac{%
d^{2}v_{x}}{d\phi ^{2}}-v_{x}Q\right) +\frac{v_{x}}{\gamma }\left( \frac{%
dv_{x}}{d\phi }\right) ^{2}-\left( 1-\frac{v_{z}}{\gamma }\right) \frac{%
dv_{x}}{d\phi }\frac{dv_{z}}{d\phi }\right]  \label{50} \\
\frac{dv_{z}}{d\phi } &=&-\frac{v_{x}}{u}AG(\phi )sin\phi +\varepsilon
\left[ u\left( \frac{d^{2}v_{z}}{d\phi ^{2}}-v_{z}Q\right) +\frac{v_{x}}{%
\gamma }\frac{dv_{x}}{d\phi }\frac{dv_{z}}{d\phi }-\left( 1-\frac{v_{z}}{%
\gamma }\right) \left( \frac{dv_{z}}{d\phi }\right) ^{2}\right]  \label{52}
\end{eqnarray}

From the above equations we derive the zeroth order for the derivatives of
the components of the acceleration

\begin{eqnarray}
\left( \frac{d^{2}v_{x}}{d\phi ^{2}}\right) _{0} &=&-A\frac{d}{d\phi }%
[G(\phi )sin\phi ]  \label{54} \\
\left( \frac{d^{2}v_{z}}{d\phi ^{2}}\right) _{0} &=&-A\frac{d}{d\phi }[\frac{%
v_{x}}{u}G(\phi )sin\phi ]  \label{55}
\end{eqnarray}

Substituting these relations back into $\left( \ref{50}\right) $ and $\left( 
\ref{52}\right) $, we obtain the first order approximation for the
components of the acceleration $\left( \frac{dv_{x}}{d\phi }\right)
_{1},\left( \frac{dv_{z}}{d\phi }\right) _{1}.$ On the other hand, the
zeroth order of the second derivative of the components of the acceleration
are given by

\begin{eqnarray}
\left( \frac{d^{3}v_{x}}{d\phi ^{3}}\right) _{0} &=&-A\frac{d^{2}}{d\phi ^{2}%
}[G(\phi )sin\phi ]  \label{56} \\
\left( \frac{d^{3}v_{z}}{d\phi ^{3}}\right) _{0} &=&-A\frac{d^{2}}{d\phi ^{2}%
}[\frac{v_{x}}{u}AG(\phi )sin\phi ]  \label{58}
\end{eqnarray}

Taking now the derivatives of $\left( \ref{50}\right) $ and $\left( \ref{52}%
\right) $ we find the rather lengthy relations

\begin{eqnarray}
\left( \frac{d^{2}v_{x}}{d\phi ^{2}}\right) _{1} &=&\left( \frac{d^{2}v_{x}}{%
d\phi ^{2}}\right) _{0}+\varepsilon \{u\left[ \left( \frac{d^{3}v_{x}}{d\phi
^{3}}\right) _{0}-v_{x}Q^{\prime }-\left( \frac{dv_{x}}{d\phi }\right)
_{1}Q\right] +u^{\prime }\left[ \left( \frac{d^{2}v_{x}}{d\phi ^{2}}\right)
_{0}-v_{x}Q\right] +  \nonumber \\
&&+2\frac{v_{x}}{\gamma }\left( \frac{dv_{x}}{d\phi }\right) _{1}\left( 
\frac{d^{2}v_{x}}{d\phi ^{2}}\right) _{0}+\frac{1}{\gamma ^{2}}\left[ \gamma
\left( \frac{dv_{x}}{d\phi }\right) _{1}-v_{x}\gamma ^{\prime }\right]
\left( \frac{dv_{x}}{d\phi }\right) _{1}^{2}-\left( 1-\frac{v_{z}}{\gamma }%
\right)   \nonumber \\
&&\left[ \left( \frac{dv_{x}}{d\phi }\right) _{1}\left( \frac{d^{2}v_{z}}{%
d\phi ^{2}}\right) _{0}+\left( \frac{dv_{z}}{d\phi }\right) _{1}\left( \frac{%
d^{2}v_{x}}{d\phi ^{2}}\right) _{0}\right] +\frac{1}{\gamma ^{2}}\left[
\gamma \left( \frac{dv_{z}}{d\phi }\right) _{1}-v_{z}\gamma ^{\prime
}\right] \left( \frac{dv_{x}}{d\phi }\right) _{1}\left( \frac{dv_{z}}{d\phi }%
\right) _{1}\}  \label{60} \\
\left( \frac{d^{2}v_{z}}{d\phi ^{2}}\right) _{1} &=&\left( \frac{d^{2}v_{z}}{%
d\phi ^{2}}\right) _{0}+\varepsilon u\left[ \left( \frac{d^{3}v_{z}}{d\phi
^{3}}\right) _{0}-v_{z}Q^{\prime }-\left( \frac{dv_{z}}{d\phi }\right)
_{1}Q\right] +u^{\prime }\left[ \left( \frac{d^{2}v_{z}}{d\phi ^{2}}\right)
_{0}-v_{z}Q\right] +  \nonumber \\
&&\frac{v_{x}}{\gamma }\left[ \left( \frac{dv_{x}}{d\phi }\right) _{1}\left( 
\frac{d^{2}v_{z}}{d\phi ^{2}}\right) _{0}+\left( \frac{dv_{z}}{d\phi }%
\right) _{1}\left( \frac{d^{2}v_{x}}{d\phi ^{2}}\right) _{0}\right] +\frac{1%
}{\gamma ^{2}}\left[ \gamma \left( \frac{dv_{x}}{d\phi }\right)
_{1}-v_{x}\gamma ^{\prime }\right] \left( \frac{dv_{x}}{d\phi }\right)
_{1}\left( \frac{dv_{z}}{d\phi }\right) _{1}  \nonumber \\
&&-2\left( 1-\frac{v_{z}}{\gamma }\right) \left( \frac{dv_{z}}{d\phi }%
\right) _{1}\left( \frac{d^{2}v_{z}}{d\phi ^{2}}\right) _{0}+\frac{1}{\gamma
^{2}}\left[ \gamma \left( \frac{dv_{z}}{d\phi }\right) _{1}-v_{z}\gamma
^{\prime }\right] \left( \frac{dv_{x}}{d\phi }\right) _{1}^{2}  \label{62}
\end{eqnarray}
where $Q^{\prime },\gamma ^{\prime }$ and $u^{\prime }$ are the derivatives
of these quantities which can be easily obtained. With these expressions
inserted  in equations $\left( \ref{50}\right) $ and $\left( \ref{52}\right)
,$ we obtain the next order approximation for the acceleration. This process
can be repeated to generate higher order corrections as desired.

The forward integration using this above procedure for the initial
acceleration was performed using the subroutine Stiffs taken from the
Numerical Recipes. The efficiency of the procedure is highly improved when
this perturbative expression of the acceleration is recurrently used at each
step of the integration process. The results of the numerical calculations
are shown in Figs. 1 and 2, respectively, for the transversal and
longitudinal components of the momentum. The parameters are $A=1$, $\epsilon
=0.05$ and $\Delta \phi =10$. These results are identical to those obtained
in Ref.\ \cite{Hart} using the backward integration method.

In conclusion, it has been shown that LD equation can be integrated forward
in time with an appropriate acceleration initial condition. This
acceleration is provided by the equation of motion itself, treating the
radiation damping term as a perturbation. This peculiar behaviour of LD
equation stems from the fact that although the self-interaction term turns
it into a third-order differential equation, only its particular solution is
physically meaningful: all homogeneous solutions have to be excluded. This
constraint makes the solution unique. We have discussed the forward
numerical integration procedure in the important case of the nonlinear
Compton scattering. The essential feature of the calculations is the crucial
role played by the recurrent use of the perturbative series expansion of the
acceleration in order to achieve stable and accurate results. We may
conclude that the final Dirac-Rohrlich constraint can indeed be replaced by
an equivalent initial condition as long as the radiation damping term can be
treated as a local perturbation in the iterative solution of the equation of
motion. This does not mean that nonlinear effects have been neglected.
Indeed, by perturbatively iterating the equation of motion, a local
expression for the acceleration is generated, in which the radiation affects
the motion as much as it is affected by it. The results presented here
clearly show that nonlinear LD equations can still be integrated {\it forward%
} in time if the appropriate, albeit unconventional, initial conditions were
utilized.

\section{Acknowledgments}

This work was partly done at Institute for Theoretical Atomic and Molecular
Physics-Harvard(ITAMP), Institute for Theoretical Physics(ITP)-Santa Barbara
and IFUSP-S\~{a}o Paulo. Partial support was supplied by the CNPq -Brazil,
FAPESP-Brazil, The National Science Foundation under Grant No. PHY94-07194
(ITP). The ITAMP is supported by the National Science Foundation.

\ {\bf Figure Captions:}

Fig. 1 The transversal component of the momentum vs. proper time, obtained
with the forward integration method. See text for details.

Fig. 2 The same for the longitudinal component of the momentum.

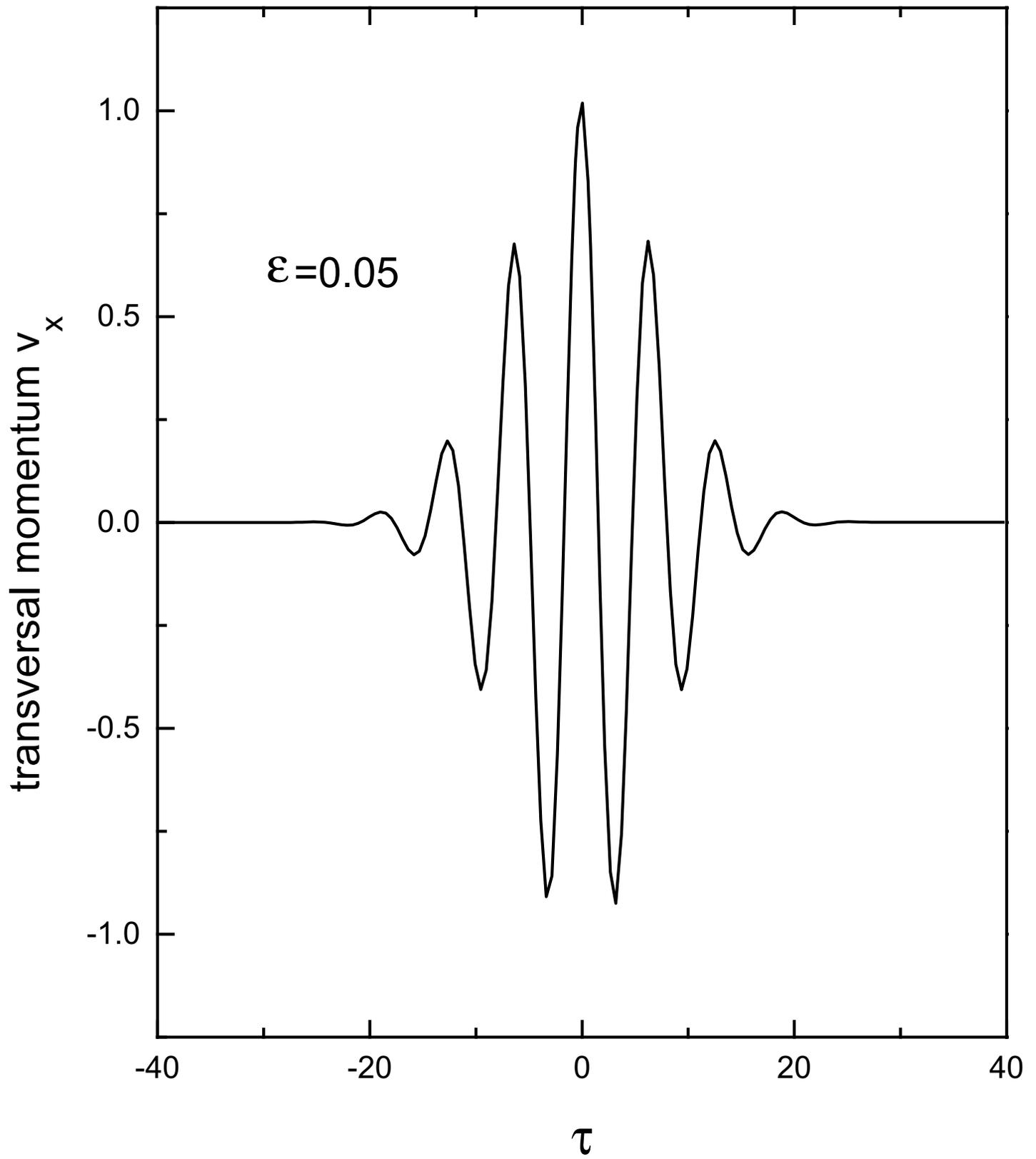
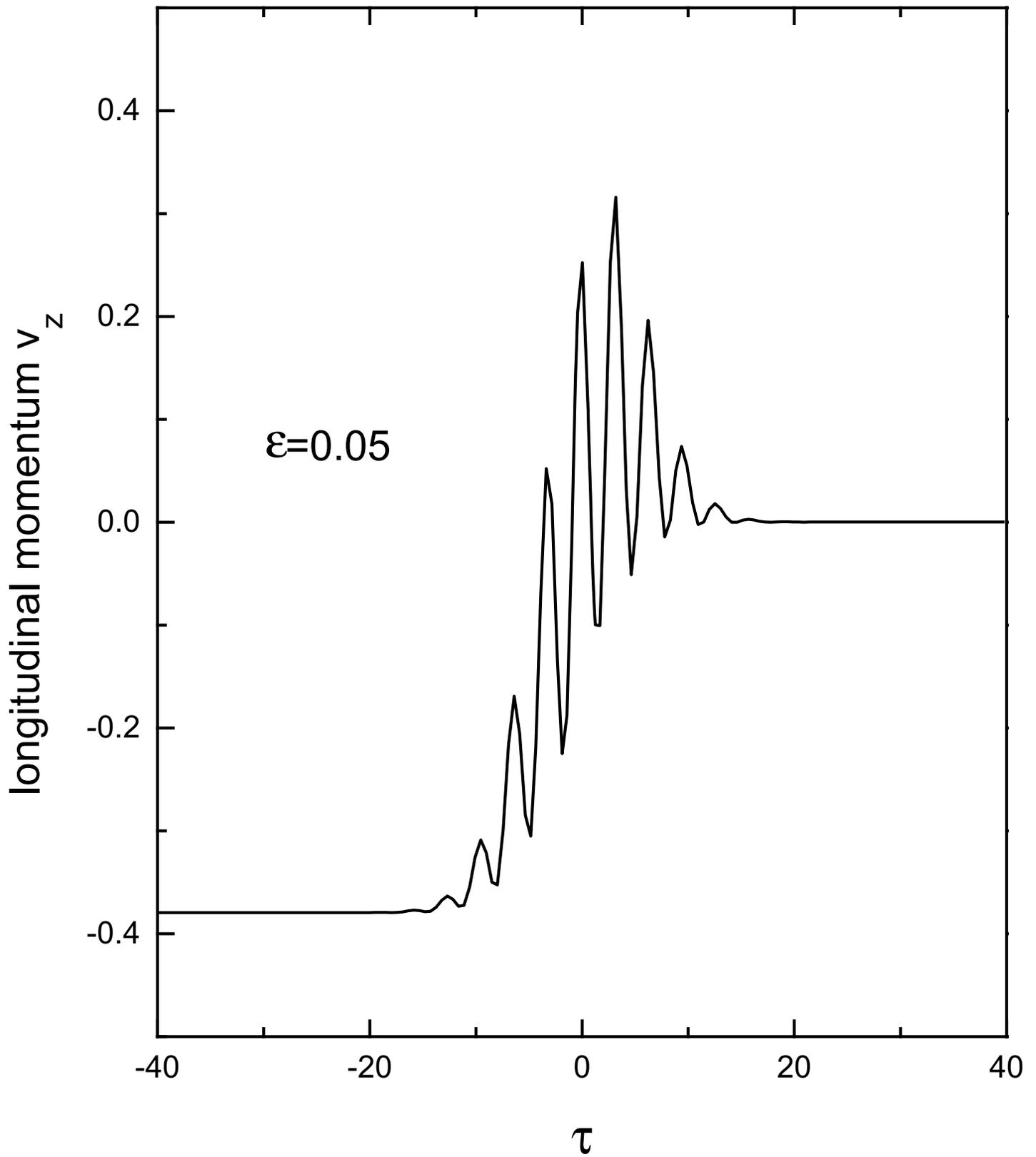
\end{document}